# High-Performance Mid-IR to Deep-UV van der Waals Photodetectors Capable of Local Spectroscopy at Room Temperature

Daozhi Shen[1,2,3], HeeBong Yang[1,4,#], Christian Spudat[1,4,5,#], Tarun Patel[1,6], Shazhou Zhong[1,6], Fangchu Chen[1,6], Jian Yan[7,8], Xuan Luo[7], Meixin Cheng[2,9], Germán Sciaini[2,9], Yuping Sun[7,10,11], Daniel A. Rhodes[12], Thomas Timusk[13], Y. Norman Zhou[3,9], Na Young Kim[1,4,9], and Adam W. Tsen[1,2,9,*]

[1] Institute for Quantum Computing, University of Waterloo, Waterloo, ON, Canada N2L 3G1
[2] Department of Chemistry, University of Waterloo, Waterloo, ON, Canada N2L 3G1
[3] Centre for Advanced Materials Joining and Department of Mechanical and Mechatronics Engineering, University of Waterloo, Waterloo, Ontario, Canada N2L 3G1
[4] Department of Electrical and Computer Engineering, University of Waterloo, Waterloo, ON, Canada N2L 3G1
[5] Institute of Electrodynamics, Microwave and Circuit Engineering, TU Wien, Gusshausstrasse 25/354, 1040 Vienna, Austria
[6] Department of Physics and Astronomy, University of Waterloo, Waterloo, ON, Canada N2L 3G1
[7] Key Laboratory of Materials Physics, Institute of Solid State Physics, HFIPS, Chinese Academy of Sciences, Hefei, China 230031
[8] University of Science and Technology of China, Hefei, China 230026
[9] Waterloo Institute for Nanotechnology, University of Waterloo, Waterloo, Ontario, Canada N2L 3G1
[10] Anhui Province Key Laboratory of Condensed Matter Physics at Extreme Conditions, High Magnetic Field Laboratory, HFIPS, Chinese Academy of Sciences, Hefei, China 230031
[11] Collaborative Innovation Center of Advanced Microstructures, Nanjing University, Nanjing, China 210093
[12] Department of Materials Science and Engineering, University of Wisconsin-Madison, WI, USA
[13] Department of Physics and Astronomy, McMaster University, Hamilton, Ontario, Canada L8S 4M1
[#] These authors contribute equally.
*Corresponding author: awtsen@uwaterloo.ca

**The ability to perform broadband optical spectroscopy with sub-diffraction-limit resolution is highly sought-after for a wide range of critical applications. However, sophisticated tip-enhanced techniques are currently required to achieve this goal. We bypass this challenge by demonstrating an extremely broadband photodetector based on a two-dimensional (2D) van der Waals heterostructure that is sensitive to light across over a decade in energy from the mid-infrared (MIR) to deep-ultraviolet (DUV) at room temperature. The devices feature high detectivity (> $10^9$ cm Hz$^{1/2}$ W$^{-1}$) together with high bandwidth (2.1 MHz). The active area can be further miniaturized to submicron dimensions, far below the diffraction limit for the longest detectable wavelength of 4.1 μm, enabling such devices for facile measurements of local optical properties on atomic-layer-thickness samples placed in close proximity. This work can lead to the development of low-cost and high-throughput photosensors for hyperspectral imaging at the nanoscale.**

Broadband optical spectroscopies, such as reflectance and absorption, from the infrared (IR) to ultraviolet (UV) are a powerful, noninvasive probe commonly used to characterize the structural, chemical, and electronic properties of all material systems[1]. Optical measurements are generally limited by diffraction, however, which imposes a tradeoff between the smallest detectable photon energy and probe area that is given by the relation: illumination spot size ≳ λ/2, at a given wavelength λ in air. This makes local spectroscopy in the longer wavelength IR regime especially difficult and imposes stringent restrictions on material uniformity that is often unachievable. One solution to this challenge is to exploit tip-enhanced near-field techniques (via either local illumination the sample and/or local light detection using a scanning subwavelength optical probe)[2]; although, they collectively carry disadvantages of high cost and low throughput. Furthermore, near-field illumination requires the use of a broadband source with sufficient intensity[3,4], while near-field detection typically suffers from low collection efficiency. Alternatively, it may be possible to obtain local spectroscopic information on nonuniform systems with conventional widefield illumination and detection by instead placing the area of interest in proximal contact to an ultrasmall photodetector. This conceptual photodetection scheme is illustrated in Figure 1a. Wavelength-dependent photocurrent measurements taken with and without the sample can then be compared to determine the local absorbance spectrum, as shown in Figure 1b. Here, the photocurrent ($I_{pc} = I_{light} - I_{dark}$) is defined as difference in device current with light ($I_{light}$) and without ($I_{dark}$).

While the latter detection scheme can be more easily implemented compared to tip-enhanced illumination, it also places stricter requirements on the capabilities of the detector: broadband response, high sensitivity, fast response speed, and miniaturizability are needed, ideally together with room temperature operability. To the best of our knowledge, such high-performance broadband detectors have not yet been demonstrated. Although thermopile sensors and some phototransistors based on 2D materials have high sensitivity and wide detection range[5–8], their response speed (several to hundreds of milliseconds) is too slow to yield spectroscopic data with reasonable throughput. Miniaturized broadband detectors present additional difficulties. Even though silicon-based sensors can be readily scaled to the micron size, they are transparent to photon energies below ~ 1.1 eV. Similarly, local photodetectors based on various nanostructures have only demonstrated a responsivity in the visible range[9,10].

To address these challenges, we first report a vertical photodetector device concept based on 2D materials with multiple bandgaps that is sensitive to radiation across a decade in energy from the mid-IR (MIR) to deep-UV (DUV), or 0.3 to 5.0 eV, at room temperature. The photocurrent spectrum is further tunable with bias voltage and can be optimized to reach a peak specific detectivity of $3.4 \times 10^9$ cm Hz$^{1/2}$ W$^{-1}$ in the MIR. The device has a high bandwidth of 2.1 MHz, the fastest yet reported for detectors with DUV to MIR sensitivity[5–7,11–13]. To understand the overall device behavior, we build a numerical finite element analysis model of the entire heterostructure and identify the bias-voltage-dependent band diagrams and carrier concentrations. Then, we show using scanning photocurrent measurements that the active area can be made far smaller than the diffraction limit for the lowest energy of detectable IR radiation, enabling such devices for direct measurements of local optical properties. We demonstrate this latter capability experimentally by performing broadband absorption spectroscopy of different 2D materials down to the monolayer limit using conventional wide-field illumination.

Figure 1a also shows a schematic of our device geometry and measurement circuit. Few-layer graphene (Gr) provides a transparent top electrode and semiconducting 2H-MoTe$_2$ is used as a middle layer to absorb light above 0.9 eV energy. Small-bandgap black phosphorus (BP) serves both as a bottom conductor and absorber for light energy down to 0.3 eV. The heterostructures are assembled by dry-transfer within a nitrogen-filled glovebox and covered by a thin insulating layer of hexagonal boron nitride (hBN). Absorbance spectra of the two semiconductors are plotted in the upper panel of Figure 2a. While BP has been previously demonstrated to be an excellent photodetector in the IR range[14–16], we find this particular combination of materials and device geometry allows for a continuous spectral response from the MIR to the DUV with exceptional performance. Substituting MoTe$_2$ with a larger bandgap semiconductor may leave a gap in the response spectrum. Although any MoTe$_2$ and BP flake thickness larger than ~ 2 nm and ~ 10 nm, respectively, will yield a bandgap close to that of the bulk crystal[17,18], we determined that specifically ~ 10-nm-thick MoTe$_2$ and ~ 17-nm-thick BP will produce higher responsivity (Figure S9).

The middle panel of Figure 2a shows the photocurrent spectral response of a typical device with relatively large junction area (15 µm × 18 µm) at several different voltage biases (see Supporting Information Section 2 for details). The full bias dependence of the spectrum is provided in Figure S3. At zero bias, the overall photoresponse is relatively weak. Upon increasing positive voltage between BP and Gr, the response in the near-IR (NIR) and visible is continuously increased, but no response is observed below 0.95 eV. The two peaks at 1.06 eV and 1.72 eV correspond to the A and A' excitons in MoTe$_2$, as can be seen in the absorption spectrum[18]. Upon increasing negative voltage, the response from MoTe$_2$ is similarly increased; however, the response from BP at lower energies increases as well. In particular, the MoTe$_2$ A exciton peak picks up where the BP response decays. The net result is a continuous spectral response from the MIR at 0.3 eV to the visible when a small negative voltage is applied to our photodetector, a key finding of this work. The decay at high energies is attributed to the low intensity of the tungsten source. As we shall show, the detector exhibits significant responsivity up to 5.0 eV, the highest energy measured, yielding a spectral range of over a full decade. For comparison, the bottom panel of Figure 2a presents the spectral response of three widely used commercial photodetectors based on Si, InGaAs and mercury cadmium telluride (MCT), respectively. Clearly, a combination of these detectors is needed to cover the same spectral range as our devices.

We now proceed to characterize the transport properties of our photodetectors in detail. Without illumination, the *I-V* characteristics of the 15 µm × 18 µm device exhibit diode-like rectifying behavior, as shown in Figure 2b. Using finite element numerical simulations of the heterostructure, we can achieve close fitting between the calculated and experimentally observed dark transport characteristics (Figure S13). A series of band alignments and carrier concentration profiles for various bias conditions are then extracted and given in Figures S14 and S15, respectively. At zero and forward bias ($V \geq 0$), the BP/MoTe$_2$ interface forms a hole-accumulation region owing to the intrinsically p-doped BP[19]. These holes are the dominant carrier concentration of the entire heterostructure, and so contribute to relatively large current levels for $V > 0$ after surmounting the MoTe$_2$ barrier. Under reverse bias ($V \lesssim -0.3V$), however, a comparably smaller population of electrons accumulate at the same interface, leading to relatively less current. This asymmetry can be directly observed in the representative band diagrams shown in Figure 2c for $V = \pm\, 0.5$ V, and is further schematized by the dark carriers and arrows in gray.

Figure 2b also shows the *I-V* characteristics of the same device under illumination by focused lasers of several different wavelengths with the same power and polarization. When we use a visible ($\lambda = 520$ nm) or NIR ($\lambda = 820$ nm) laser, whose energy is larger than the MoTe$_2$ bandgap energy, we observe a large increase in device current with light, or photocurrent $I_{pc}$, for both bias directions. As photocarriers are predominantly generated in the upper MoTe$_2$ layer for this photon energy regime, both forward and reverse bias lead to barrierless transport and relatively efficient collection of these carriers, and so $I_{pc} > I_{dark}$ for nearly all *V*. This effect is captured by the blue carriers and arrows in Figure 2c.

When we use IR lasers ($\lambda = 1310, 2400$ nm), on the other hand, large photocurrent flows only under reverse bias, and not forward bias, which is consistent with the bias dependence of the full photocurrent spectrum (Figure 2a). In general, we expect the photocurrent to be suppressed if the number of dark carriers greatly exceed that of the photo-generated carriers. As the interfacial carrier concentration under forward bias is significantly greater than that under reverse bias (Figure S15), $I_{pc} \ll I_{dark}$ for $V \geq 0$ when the photon energy is below the MoTe$_2$ bandgap. This bias asymmetry is illustrated by the red carriers and arrows in Figure 2c. In short, to obtain the broadest spectral photoresponse, BP must be negatively biased against Gr in our device geometry, and so we primarily focus on this regime for the remainder of this work.

We next set out to determine the optimal light excitation and bias conditions at which to evaluate the various performance metrics of our 15 µm × 18 µm photodetector (responsivity, quantum efficiency, detectivity, and response speed). First, in Figure 3a, we show the laser power dependence of $I_{pc}$ for various incident wavelengths. The response is linear below ~ 5 µW for all wavelengths, but gradually saturates with increasing power, and so we shall stay within the linear regime for subsequent measurements.

Figure 3b displays normalized $I_{pc}$ as a function of the linear polarization angle of the laser for three wavelengths. For $\lambda = 2300$ nm (0.52 eV), light absorption is almost solely attributed to BP, which is known to be stronger (weaker) for polarization along the armchair (zigzag) direction[14,20,21]. We thus identify the maximal $I_{pc}$ lobes as the BP armchair angle. For $\lambda = 520$ nm (2.4 eV), $I_{pc}$ is almost fully symmetric, which indicates that the photocarriers are mainly generated by light absorption from MoTe$_2$. For $\lambda = 1310$ nm (0.95 eV), we observe an intermediate level of asymmetry.

For the polarization angle yielding maximum $I_{pc}$, we show in Figure 3c the device responsivity ($\Re$) as a function of laser energy up to 3.1 eV and bias voltage in a continuous 2D false-color plot. We have boxed the regions where BP or MoTe$_2$ primarily contribute to the photoresponse. Photocurrent at high energies is mainly due to MoTe$_2$ absorption at all biases, with an optimum $\Re$ of ~ 0.15 A/W at 1.88 eV and −0.23 V. The low energy response at negative bias is due to BP and exhibits $\Re$ as high as ~ 0.2 A/W at 0.54 eV and −0.3 V. While the spectral response is highly tunable with voltage, it appears a single negative bias near −0.3 V (marked by the red arrow) can be applied to reach peak responsivity for all wavelengths (Figure S7). This unique feature of our photodetectors allows us to access their full spectral range without changing biasing conditions as is required for some other detector systems[22,23]. We note that the optimal voltage is dependent on the incident light power (Figure S8).

In Figure 3d, we present the external quantum efficiency (EQE = $\Re hc/(\lambda e)$, where $h$ is Planck's constant, $c$ is speed of light in vacuum, and $e$ is the electron charge) and specific detectivity ($D^* = \Re A^{1/2}/S_n$, where $A$ is the detector area and $S_n$ is the current noise spectral density averaged over the electrical bandwidth), obtained at −0.3 V as a function of laser energy up to 5.0 eV (see Supporting Information Section 4 for details of noise measurements). The frequency-dependent noise performance for this and another smaller device is shown in Figure S12. The EQE peaks at 30% in the near UV (3.1 eV), while the highest $D^* = 3.4 \times 10^9$ cm Hz$^{1/2}$ W$^{-1}$ is observed in the MIR (0.52 eV), which is comparable to that of commercial IR detectors based on InAsSb or InSb[24]. Furthermore, $D^*$ remains on the order of $10^9$ cm Hz$^{1/2}$ W$^{-1}$ throughout the entire spectral range (except at the highest energy measured at 5.0 eV), enabling low light detection from MIR to all the way to DUV.

To determine the speed of the device, we measured its frequency response to obtain the −3 dB bandwidth. The upper panel of figure 3e shows time-domain photocurrent under irradiation by the 658 nm laser with a 2 MHz sinusoidal intensity modulation. The clear oscillations indicate that the device is responsive even at this high frequency. The frequency dependence of the normalized photocurrent amplitude is shown in the bottom panel of Figure 3e, from which the –3dB bandwidth is determined to be 2.1 MHz. This speed is the fastest reported for photodetectors with comparable spectral range[5–7,11–13]. For comparison, we have plotted the electrical bandwidth versus wavelength range for various commercially available and reported broadband detectors in Figure 3f. Our detectors have a clear performance advantage as they can achieve ultrafast response speed together with extremely broadband detection without the need for cryogenic cooling.

A key advantage of our device geometry is that the junction area size can be easily controlled to be smaller than the diffraction limit for IR radiation, enabling it for local spectroscopic measurements. First, we need to confirm that the active region is indeed localized within the overlap area between the three materials. We have thus performed spatially resolved photocurrent measurements by raster-scanning the focused laser. Figure 4 shows photocurrent imaging of two devices with different junction sizes (16 µm × 17 µm and 0.8 µm × 0.8 µm), each taken using two laser wavelengths in the visible (658 nm) and IR (2300 nm). The laser reflection images are shown in the panels on the left in grayscale. For the 658 nm laser, the spot size formed by the objective lens is ~ 2 µm (Figure S6a), which is larger than the junction area of the smaller device. Nonetheless, the strongest signal is clearly detected to originate from the overlap region of both, although there is a small decay outside the junction along the BP. The small features observed in the junction of the larger device correspond to unintended nonuniformities (wrinkles, bubbles, etc.) created during the fabrication process and can be recognized in the optical image as well.

For the 2300 nm laser, photocurrent images are taken using a reflective objective, which produces a series of circular fringes around the main focal spot (Figure S6b). We thus attribute the bright features outside of the junction to imaging artifacts and not to the detectors themselves. We have further taken photocurrent spectra of both devices using the FTIR (Figure S4). While the MIR response is slightly reduced in the device with smaller area, both produce a substantial photocurrent response down to 0.3 eV. These results establish that our photodetector junction can still be responsive even when the size is scaled five times below its longest detectable wavelength of 4.1 µm.

We now finally demonstrate experimentally that our sub-wavelength photodetectors are sensitive enough to perform absorption spectroscopy on samples locally across the entire spectral range of detectivity. We have fabricated several small-area devices with junction sizes between 0.8 µm × 0.8 µm and 3 µm × 3 µm and illuminated them using a slightly focused tungsten lamp, which forms a relatively large ~ 3 mm diameter spot centered on the junction. While it may be possible to position 2D materials stably and closely on top using piezos for absorbance measurements, we opt to directly transfer the flakes as a simple proof-of-principle. The inset of Figure 5a shows an optical image of a representative detector with 15-nm-thick $Ta_2NiSe_5$ above the active area and the upper panel shows photocurrent spectra in log scale taken before and after the transfer. The difference between the two curves is shaded in gray and yields the local absorbance spectrum shown in the lower panel. The α and β peaks at 1.5 eV and 2.2 eV, respectively, have been previously identified in bulk crystals[25]. Importantly, we are also able to detect the peak in the MIR at ~ 0.39 eV (3180 nm)[26].

We have similarly obtained the absorbance spectrum of monolayer $WSe_2$ using this technique. An optical image of another detector with $WSe_2$ above is shown in the inset of Figure 5b. The color of the $WSe_2$ has been enhanced to highlight the thickness variation across the flake. Despite the sample nonuniformity, we can perform local spectroscopy only on the monolayer aligned on top of the junction. Even though the difference between the photocurrent spectra with and without $WSe_2$ is small (Figure 5b, upper panel), the absorbance spectrum can still be clearly extracted (Figure 5b, lower panel). The low overall sample absorbance (~ 0.06 maximum) is consistent with that of previous measurements using conventional techniques[27]. The direct bandgap absorption onset edge is clearly seen at ~ 1.5 eV, and the A and B excitons at 1.67 eV and 2.1 eV can be identified as well, all in agreement with the expected characteristics of monolayer $WSe_2$[28]. Our photodetectors are thus able to resolve absorption features locally down to the single layer limit.

In summary, we have demonstrated a new photodetector concept based on 2D van der Waals heterostructures with high broadband detectivity from the MIR to DUV and fast response times that is operable at room temperature. The active device area can be scaled down to 0.6 µm$^2$ and still sense IR radiation with wavelength longer than 4 µm. Furthermore, the sensitivity is high enough to perform absorption spectroscopy on monolayer flakes under conventional widefield illumination. Such devices provide a new low-cost route towards local IR measurements beyond the diffraction limit without the need for high-cost and restrictive near-field techniques. By replacing BP with black arsenic phosphorus, the longest detectable wavelength may potentially be expanded to 8.2 µm[29]. Using nanofabrication, one may be able to decrease the size of the junction further. Combined with large-scale films grown by chemical vapor deposition[30], one may even be able to develop dense pixel arrays using our heterostructure to perform hyperspectral imaging with super-resolution across over a decade in energy in the future.

**Acknowledgements**

We thank Michael Reimer and Jayakanth Ravichandran for a critical reading of our manuscript, and thank P. Sprenger and F. Sfigakis for assistance with experiments. A.W.T. acknowledges support from the US Army Research Office (W911NF-21-2-0136), Ontario Early Researcher Award (ER17-13-199), and the National Science and Engineering Research Council of Canada (NSERC) (RGPIN-2017-03815). This research was undertaken thanks in part to funding from the Canada First Research Excellence Fund. H.B.Y. and N.Y.K. acknowledge the support of Industry Canada and the Ontario Ministry of Research & Innovation through Early Researcher Awards (RE09-068). H.B.Y., C.S., and N.Y.K. acknowledge CMC Microsystems for COMSOL Multiphysics. Y.N.Z. and D.S. acknowledge NSERC and the Canada Research Chairs Program. T.T. was supported by NSERC grant "Optics of quantum materials." D.A.R. acknowledges the support provided by the University of Wisconsin-Madison Office of the Vice Chancellor for Research and Graduate Education with funding from the Wisconsin Alumni Research Foundation. J.Y., X.L. and Y.P.S. thank the financial support from the National Nature Science Foundation of China under Contract Nos. 11674326, 11874357, the Joint Funds of the National Natural Science Foundation of China, the Chinese Academy of Sciences' Large-Scale Scientific Facility under Contract Nos. U1832141, U1932217, and U2032215, the Key Research Program of Frontier Sciences, CAS (No. QYZDB-SSW-SLH015), the uses with Excellence and Scientific Research Grant of Hefei Science Center of CAS (No.2018HSC-UE011).

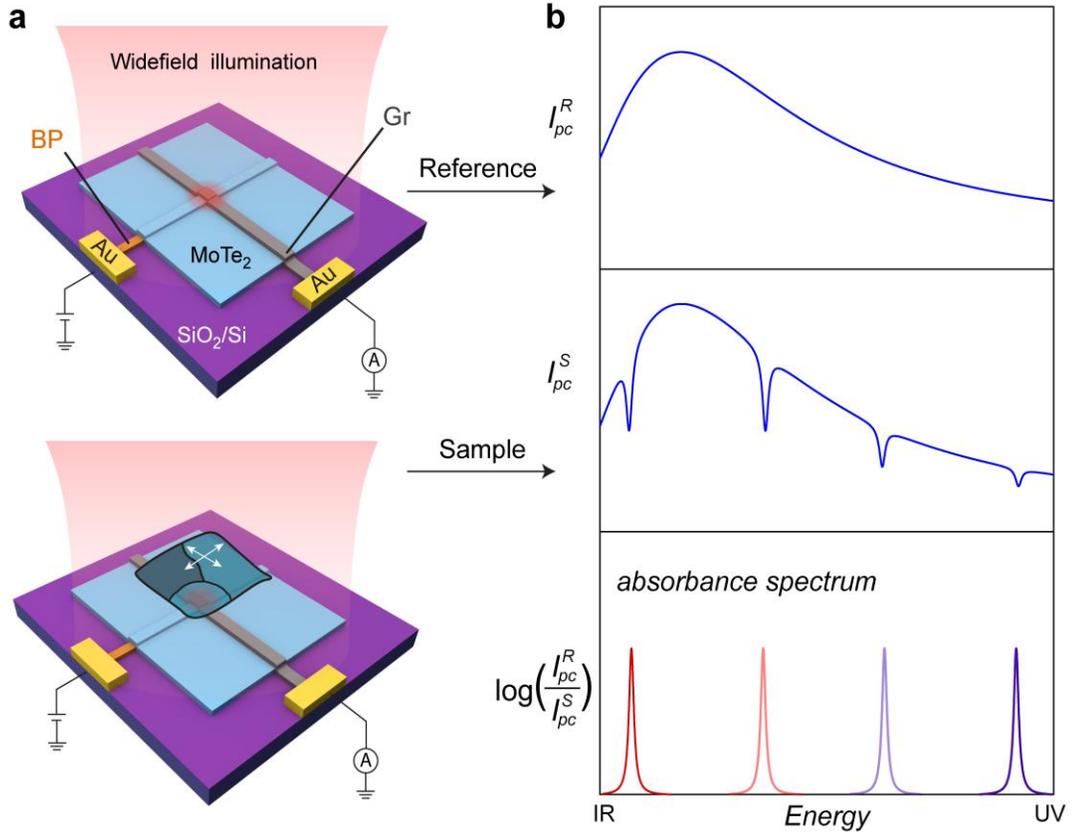

**Figure 1. Miniature broadband photodetector based on 2D materials for local spectroscopy using widefield illumination. a**, Geometry and measurement schematics of vertical BP/MoTe$_2$/Gr device before (top) and after (bottom) a nonuniform sample is aligned above the detector area (movement symbolized by white arrows). hBN covering layer is not shown. **b**, Corresponding photocurrent spectra with ($I_{pc}^S$) and without ($I_{pc}^R$) the sample. The local absorbance spectrum of the sample is obtained from logarithmic ratio of the two photocurrents, $\log(I_{pc}^R/I_{pc}^S)$.

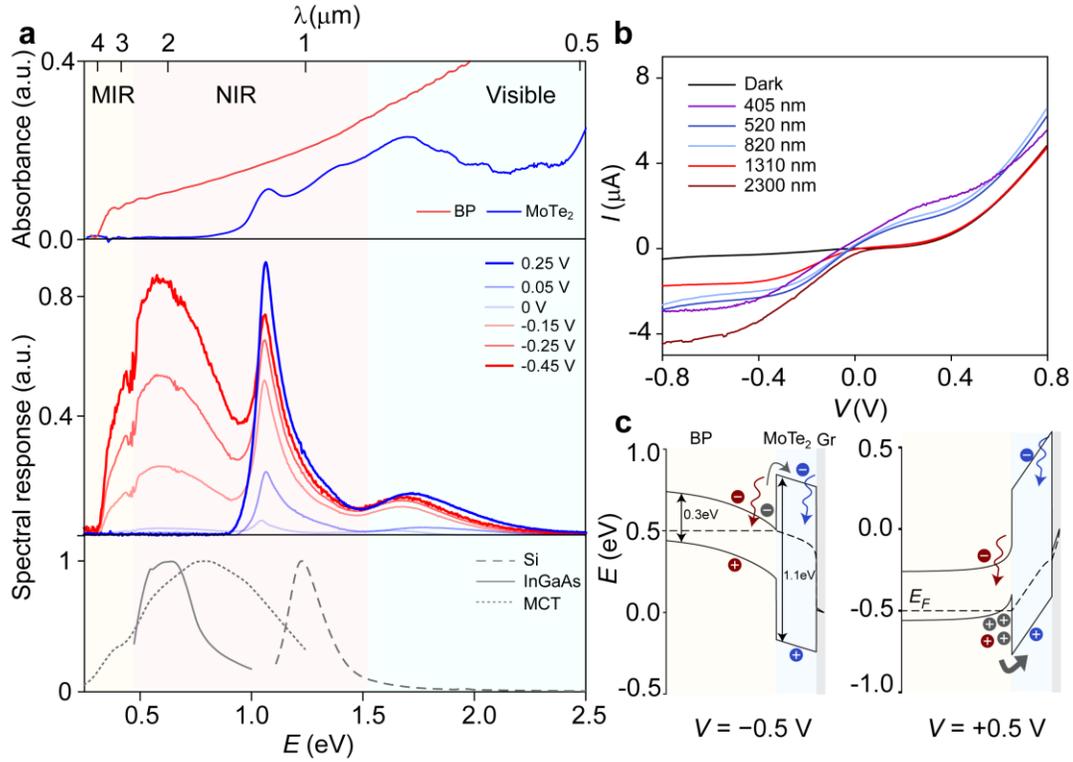

**Figure 2. Spectral response and device modeling. a,** (Top) Absorbance spectra of BP and MoTe$_2$. (Middle) Photocurrent spectra at six bias voltages taken under illumination by a broadband tungsten lamp. (Bottom) The light source spectra measured using three widely used commercial photodetectors based on Si, InGaAs, and MCT. **b,** *I-V* curves of device in dark and under illumination by focused lasers of several different wavelengths with 30 µW power and linear polarization fixed in direction generating maximum photocurrent. No substantial photocurrent is observed for IR light at positive *V*. **c,** Band structure diagrams of the heterostructure for *V* = −0.5 V (top) and *V* = +0.5 V (bottom) extracted from finite element simulations. Dark carriers in BP near the MoTe$_2$ interface are shown in gray. Photocarriers generated by IR and visible light are colored in red and blue, respectively. The large interfacial BP hole population under forward bias yields insignificant IR $I_{pc}$ relative to $I_{dark}$.

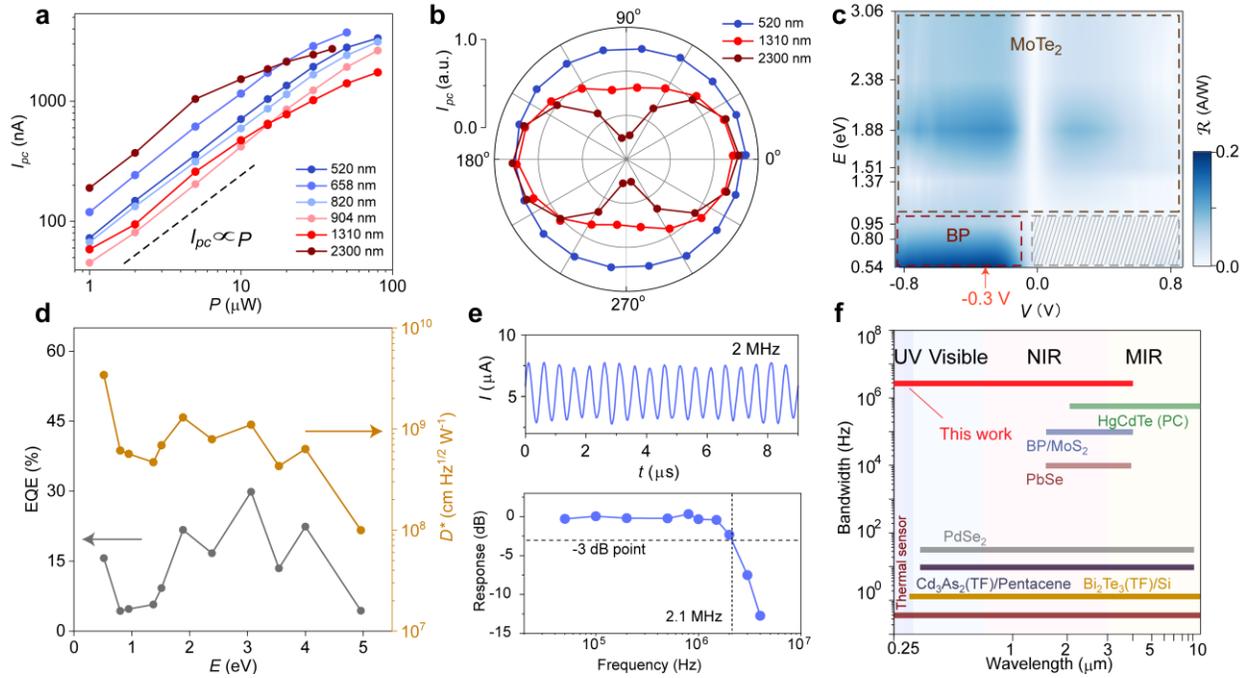

**Figure 3. Characterization of photodetector performance metrics.** Photocurrent $I_{pc}$ as function of **a,** laser power and **b,** linear polarization angle for different laser wavelengths. **c,** 2D map of detector responsivity vs bias voltage and laser energy. The different photoresponse regions are boxed by dashed lines. Red arrow marks the voltage where quantum efficiency, detectivity, and rise/fall time are evaluated. The hashed area corresponds to where $I_{pc}$ is insubstantial. **d**, External quantum efficiency (EQE) and specific detectivity ($D^*$) vs laser energy. **e,** (Upper) Photocurrent oscillations under 658 nm laser with 2 MHz sinusoidal modulation. (Bottom) Photocurrent frequency response. **f**, Electrical bandwidth vs spectral response of Gr/MoTe$_2$/BP device compared with commercially available and reported broadband photodetectors.

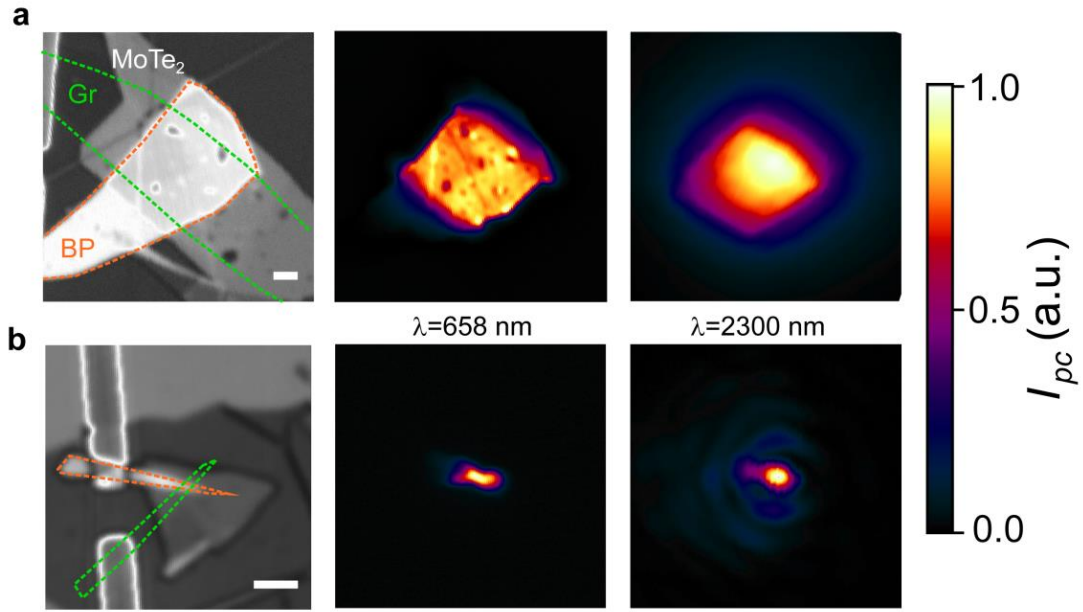

**Figure 4. Spatially resolved photocurrent imaging of detectors**. Reflection (left), 658 nm photocurrent (middle), and 2300 nm photocurrent (right) images of devices with **a,** 16 µm × 17 µm and **b,** 0.8 µm × 0.8 µm junction areas. Orange and green dashed lines outline the BP and Gr, respectively. Scale bars: 5 µm. Fringes in IR photocurrent are due to the use of a reflective objective.

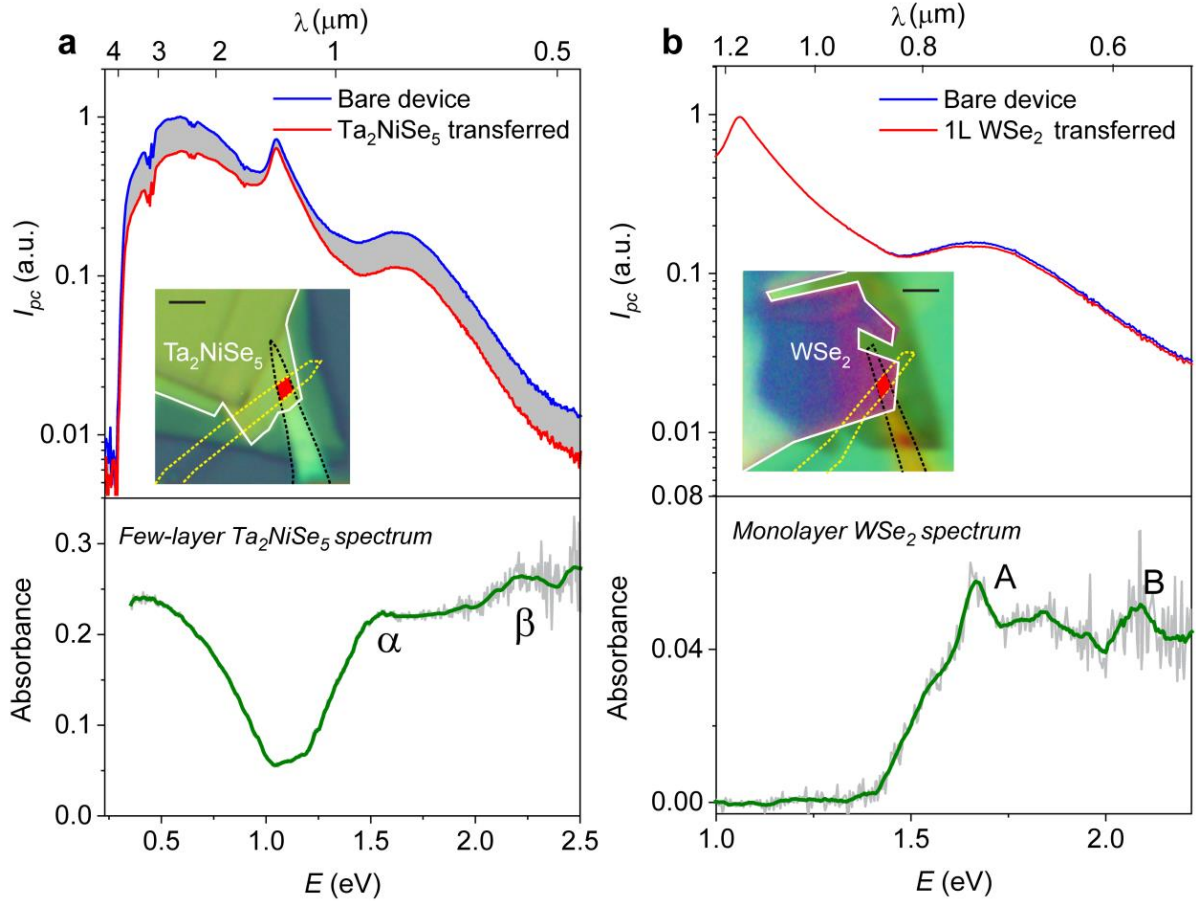

**Figure 5. Demonstration of local spectroscopy capability on non-uniform 2D samples. a,** Few-layer $Ta_2NiSe_5$ and **b,** Monolayer (1L) $WSe_2$ spectra. The inset images show ultrasmall photodetector devices (0.8 µm × 0.8 µm) after non-uniform flakes are transferred on top. Black dashed lines, yellow dashed lines, and white solid lines outline the BP, Gr, and 2D sample, respectively. Scale bars: 2 µm. A broadband tungsten source is focused by a parabolic mirror to a ~3 mm spot centered on the device junction marked by the red shaded area. The upper panels show photocurrent spectra of detectors with and without the 2D sample (difference shaded in gray). The bottom panels show absorbance spectra of thin $Ta_2NiSe_5$ and monolayer $WSe_2$ obtained from the procedure described in Figure 1. A background subtraction has been applied for $WSe_2$. Raw (smoothed) absorbance data is shown in light gray (dark green). Higher noise at higher energy is due to lower intensity of the source. The photodetector can both resolve features down to a single atomic layer and yield broadband sensitivity down to 0.3 eV or 4.1 µm, larger than the width of the junction.